\documentclass[article,shortnames,nojss]{jss}

\usepackage{
	amsfonts,
	amsmath,
	booktabs,
	bm}

\author{Quentin F. Gronau\\University of Amsterdam \And 
        Henrik Singmann\\University of Zurich \And 
        Eric-Jan Wagenmakers\\University of Amsterdam}
\title{\pkg{bridgesampling}: An \proglang{R} Package for Estimating Normalizing Constants}

\Plainauthor{Quentin F. Gronau, Henrik Singmann, Eric-Jan Wagenmakers} 
\Plaintitle{bridgesampling: An R Package for Estimating Normalizing Constants} 
\Shorttitle{\pkg{bridgesampling}} 

\Abstract{Statistical procedures such as Bayes factor model selection and Bayesian model averaging require the computation of normalizing constants (e.g., marginal likelihoods). These normalizing constants are notoriously difficult to obtain, as they usually involve high-dimensional integrals that cannot be solved analytically. Here we introduce an \proglang{R} package that uses bridge sampling \citep{MengWong1996,MengSchilling2002} to estimate normalizing constants in a generic and easy-to-use fashion. For models implemented in \proglang{Stan}, the estimation procedure is automatic. We illustrate the functionality of the package with three examples.
}
\Keywords{bridge sampling, marginal likelihood, model selection, Bayes factor, Warp-III}
\Plainkeywords{bridge sampling, normalizing constant, model selection, Bayes factor, Warp-III} 

\Address{
  Quentin F. Gronau\\
  Department of Psychological Methods\\
  University of Amsterdam\\
  Nieuwe Achtergracht 129 B\\
  1018 WT Amsterdam, The Netherlands\\
  E-mail: \email{Quentin.F.Gronau@gmail.com}
}

\begin{document}

\section{Introduction}

In many statistical applications, it is essential to obtain normalizing constants of the form
\begin{equation}
\label{eq:normalizing_constant}
Z = \int_{\bm{\Theta}} q(\bm{\theta}) \, \text{d}\bm{\theta},
\end{equation}
where $p(\bm{\theta}) = q(\bm{\theta})/Z$ denotes a probability density function (pdf) defined on the domain $\bm{\Theta} \subseteq \mathbb{R}^p$. For instance, the estimation of normalizing constants plays a crucial role in free energy estimation in physics, missing data analyses in likelihood-based approaches, Bayes factor model comparisons, and Bayesian model averaging \citep[e.g.,][]{GelmanMeng1998}. In this article, we focus on the role of the normalizing constant in Bayesian inference; however, the \pkg{bridgesampling} package can be used in any context where one desires to estimate a normalizing constant.

In Bayesian inference, the normalizing constant of the joint posterior distribution is involved in (a) parameter estimation, where the normalizing constant ensures that the posterior integrates to one; (b) Bayes factor model comparison, where the ratio of normalizing constants quantifies the data-induced change in beliefs concerning the relative plausibility of two competing models \citep[e.g.,][]{KassRaftery1995}; (c) Bayesian model averaging, where the normalizing constant is required to obtain posterior model probabilities \citep[BMA;][]{HoetingEtAl1999}.

For Bayesian parameter estimation, the need to compute the normalizing constant can usually be circumvented by the use of sampling approaches such as Markov chain Monte Carlo \citep[MCMC; e.g.,][]{GamermanLopes2006}. However, for Bayes factor model comparison and BMA, the normalizing constant of the joint posterior distribution -- in this context usually called \emph{marginal likelihood} -- remains of essential importance. This is evident from the fact that the posterior model probability of model $\mathcal{M}_i, \, i \in\{1,2,\ldots,m\},$ given data $\bm{y}$ is obtained as
\begin{equation}
\label{eq:post_model_probs}
\underbrace{p(\mathcal{M}_i \mid \bm{y})}_{\text{posterior model probability}} = \;\;\;\;\; \underbrace{\frac{p(\bm{y} \mid \mathcal{M}_i)}{\sum_{j=1}^{m} p(\bm{y} \mid \mathcal{M}_j) \, p(\mathcal{M}_j)}}_{\text{updating factor}} \;\;\;\;\;\; \times \underbrace{p(\mathcal{M}_i)}_{\text{prior model probability}},
\end{equation}
where $p(\bm{y} \mid \mathcal{M}_i)$ denotes the \emph{marginal likelihood} of model $\mathcal{M}_i$.

If the model comparison involves only two models, $\mathcal{M}_1$ and $\mathcal{M}_2$, it is convenient to consider the odds of one model over the other. Bayes' rule yields:
\begin{equation}
\label{eq:post_model_odds}
\underbrace{\frac{p(\mathcal{M}_1 \mid \bm{y})}{p(\mathcal{M}_2 \mid \bm{y})}}_{\text{posterior odds}} = \underbrace{\frac{p(\bm{y} \mid \mathcal{M}_1)}{p(\bm{y} \mid \mathcal{M}_2)}}_{\text{Bayes factor BF$_{12}$}} \times \;\;\, \underbrace{\frac{p(\mathcal{M}_1)}{p(\mathcal{M}_2)}}_{\text{prior odds}}.
\end{equation}
The change in odds brought about by the data is given by the ratio of the marginal likelihoods of the models and is known as the \emph{Bayes factor} \citep{Jeffreys1961,KassRaftery1995,etz2017jbs}. Equation~(\ref{eq:post_model_probs}) and Equation~(\ref{eq:post_model_odds}) highlight that the normalizing constant of the joint posterior distribution, that is, the marginal likelihood, is required for computing both posterior model probabilities and Bayes factors.

The marginal likelihood is obtained by integrating out the model parameters with respect to their prior distribution:
\begin{equation}
\label{eq:marginal_likelihood}
p(\bm{y} \mid \mathcal{M}_i) = \int_{\bm{\Theta}} p(\bm{y} \mid \bm{\theta}, \mathcal{M}_i) \thinspace p(\bm{\theta} \mid \mathcal{M}_i) \, \text{d}\bm{\theta}.
\end{equation}
The marginal likelihood implements the principle of parsimony also known as \emph{Occam's razor} \citep[e.g.,][]{JefferysBerger1992, MyungPitt1997,VandekerckhoveEtAl2015MPT}.
Unfortunately, the marginal likelihood can be computed analytically for only a limited number of models. For more complicated models (e.g., hierarchical models), the marginal likelihood is a high-dimensional
integral that usually cannot be solved analytically.
This computational hurdle has complicated the application of Bayesian model comparisons for decades.

To overcome this hurdle, a range of different methods have been developed that vary in accuracy, speed, and complexity of implementation: naive Monte Carlo estimation, importance sampling, the generalized harmonic mean estimator, Reversible Jump MCMC \citep{Green1995}, the product-space method \citep{CarlinChib1995,LodewyckxEtAl2011}, Chib's method \citep{Chib1995}, thermodynamic integration \citep[e.g.,][]{LartillotPhilippe2006}, path sampling \citep{GelmanMeng1998}, and others.
The ideal method is fast, accurate, easy to implement, general, and unsupervised, allowing non-expert users to treat it as a ``black box''.

In our experience, one of the most promising methods for estimating normalizing constants is bridge sampling \citep{MengWong1996,MengSchilling2002}. Bridge sampling is a general procedure that performs accurately even in high-dimensional parameter spaces such as those that are regularly encountered in hierarchical models. In fact, simpler estimators such as the naive Monte Carlo estimator, the generalized harmonic mean estimator, and importance sampling are special sub-optimal cases of the bridge identity described in more detail below \citep[e.g.,][]{FruehwirthSchnatter2004,GronauEtAltutorial}.

In this article, we introduce \pkg{bridgesampling}, an \proglang{R} \citep{R} package that enables the straightforward and user-friendly estimation of the marginal likelihood (and of normalizing constants more generally) via bridge sampling techniques. In general, the user needs to provide to the \code{bridge_sampler} function four quantities that are readily available:
\begin{itemize}
\item an object with posterior samples (argument \code{samples});
\item a function that computes the log of the unnormalized posterior density for a set of model parameters (argument \code{log_posterior});
\item a data object that contains the data and potentially other relevant quantities for evaluating \code{log_posterior} (argument \code{data});
\item lower and upper bounds for the parameters (arguments \code{lb} and \code{ub}, respectively).
\end{itemize}
Given these inputs, the \pkg{bridgesampling} package provides an estimate of the log marginal likelihood.

\begin{figure}[!tb]
 \begin{center}
	\includegraphics[width = .8\textwidth]{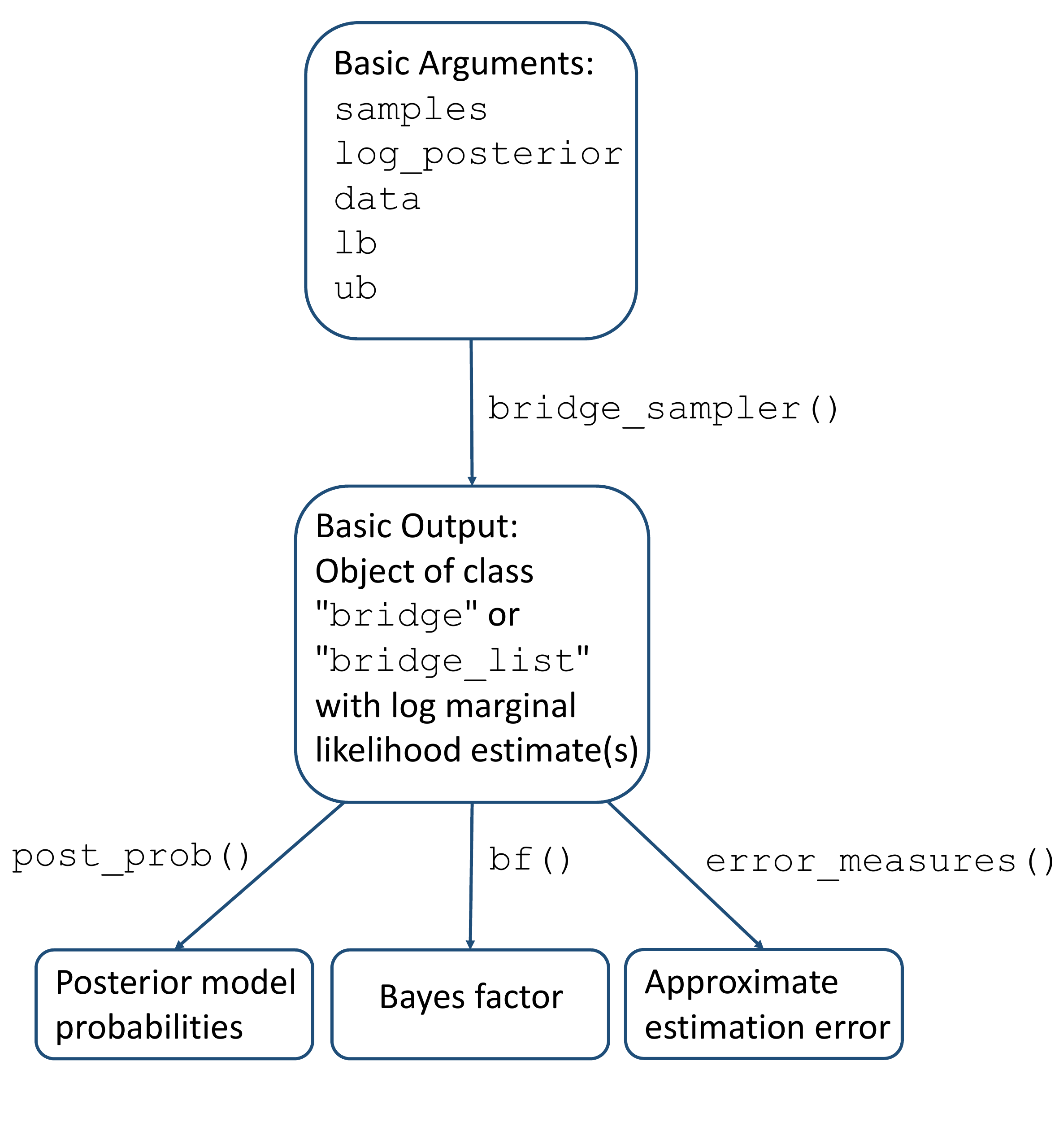}
	\caption{Flow chart of the steps that a user may take when using the \pkg{bridgesampling} package. In general, the user needs to provide a posterior samples object (\code{samples}), a function that computes the log of the unnormalized posterior density (\code{log\_posterior}), the data (\code{data}), and parameter bounds (\code{lb} and \code{ub}). The \code{bridge\_sampler} function then produces an estimate of the log marginal likelihood. This is usually repeated for at least two different models. The user can then compute posterior model probabilities (using the \code{post\_prob} function), Bayes factors (using the \code{bf} function), and approximate estimation errors (using the \code{error\_measures} function). Note that the summary method for bridge objects automatically invokes the \code{error\_measures} function. Figure available at \url{https://tinyurl.com/ybf4jxka} under CC license \url{https://creativecommons.org/licenses/by/2.0/}.}
	\label{fig:flow_chart}
	 \end{center}
\end{figure}

Figure~\ref{fig:flow_chart} displays the steps that a user may take when using the \pkg{bridgesampling} package. Starting from the top, the user provides the basic required arguments to the \code{bridge_sampler} function which then produces an estimate of the log marginal likelihood. With this estimate in hand -- usually for at least two different models -- the user can compute posterior model probabilities using the \code{post_prob} function, Bayes factors using the \code{bf} function, and approximate estimation errors using the \code{error_measures} function.
A schematic call of the \code{bridge_sampler} function looks as follows (detailed examples are provided in the next sections):
\begin{Sinput}
R> bridge_sampler(samples = samples, log_posterior = log_posterior,
+                 data = data, lb = lb, ub = ub)
\end{Sinput}

The \code{bridge_sampler} function is an \code{S3} generic which currently has methods for objects of class \code{mcmc}, \code{mcmc.list} \citep{PlummerEtAl2006}, \code{stanfit} \citep{rstan}, \code{matrix}, \code{rjags} \citep{rjags,R2jags}, \code{runjags} \citep{runjags}, \code{stanreg} \citep{rstanarm}, and for \code{MCMC\_refClass} objects produced by \pkg{nimble} \citep{nimble2017}.\footnote{We thank Ben Goodrich for adding the \code{stanreg} method to our package and Perry de Valpine for his help implementing the \pkg{nimble} support.}
This allows the user to obtain posterior samples in a convenient and efficient way, for instance, via \proglang{JAGS} \citep{Plummer2003} or a highly customized sampler. Hence, bridge sampling does not require users to program their own MCMC routines to obtain posterior samples; this convenience is usually missing for methods such as Reversible Jump MCMC \citep[but see][]{rjmcmc}.

When the model is specified in \proglang{Stan} \citep{CarpenterEtAl2017, rstan} -- in a way that retains the constants, as described below -- obtaining the marginal likelihood is even simpler: the user only needs to pass the \code{stanfit} object to the \code{bridge_sampler} function.
The combination of \proglang{Stan} and the \pkg{bridgesampling} package therefore produces an unsupervised, black box computation of the marginal likelihood.

This article is structured as follows: First we describe the implementation details of the algorithm from \pkg{bridgesampling}; second, we illustrate the functionality of the package using a simple Bayesian $t$-test example where posterior samples are obtained via \proglang{JAGS}.
In this section, we also explain a heuristic to obtain the function that computes the log of the unnormalized posterior density in \proglang{JAGS}; third, we describe in more detail the interface to \proglang{Stan} which enables an even more automatized computation of the marginal likelihood. Fourth, we illustrate use of the \proglang{Stan} interface with two well-known examples from the Bayesian model selection literature.

\section{Bridge sampling: the algorithm}
Bridge sampling can be thought of as a generalization of simpler methods for estimating normalizing constants such as the naive Monte Carlo estimator, the generalized harmonic mean estimator, and importance sampling \citep[e.g.,][]{FruehwirthSchnatter2004,GronauEtAltutorial}.
These simpler methods typically use samples from a single distribution, whereas bridge sampling combines samples from \emph{two} distributions.\footnote{Note, however, that these simpler methods are special cases of bridge sampling \citep[e.g.,][Appendix A]{GronauEtAltutorial}. Hence, for particular choices of the bridge function and the proposal distribution, only samples from one distribution are used.}
For instance, in its original formulation \citep{MengWong1996}, bridge sampling was used to estimate a ratio of two normalizing constants such as the Bayes factor. In this scenario, the two distributions for the bridge sampler are the posteriors for each of the two models involved.
However, the accuracy of the estimator depends crucially on the overlap between the two involved distributions; consequently, the accuracy can be increased by estimating a single normalizing constant at a time, using as a second distribution a convenient normalized proposal distribution that closely matches the distribution of interest \citep[e.g.,][]{GronauEtAltutorial,OverstallForster2010}. The bridge sampling estimator of the marginal likelihood is then given by:\footnote{We omit conditioning on the model for enhanced legibility. It should be kept in mind, however, that this yields the estimate of the marginal likelihood for a particular model $\mathcal{M}_i$, that is, $p(\bm{y} \mid \mathcal{M}_i)$.}
\begin{equation}
\label{eq:bridge_identity}
p(\bm{y}) = \frac{\mathbb{E}_{g(\bm{\theta})}\left[h(\bm{\theta}) \, p(\bm{y} \mid \bm{\theta}) \, p(\bm{\theta})\right]}{\mathbb{E}_{p(\bm{\theta} \mid \bm{y})}\left[h(\bm{\theta}) \, g(\bm{\theta})\right]} \approx \frac{\frac{1}{n_2}\sum_{j=1}^{n_2} h(\tilde{\bm{\theta}}_j) \, p(\bm{y} \mid \tilde{\bm{\theta}}_j) \, p(\tilde{\bm{\theta}}_j)}{\frac{1}{n_1}\sum_{i = 1}^{n_1}h(\bm{\theta}^\ast_i) \, g(\bm{\theta}^\ast_i)},
\end{equation}
where $h(\bm{\theta})$ is called the \textit{bridge function} and $g(\bm{\theta})$ denotes the \textit{proposal distribution}. \sloppy$\{\bm{\theta}^\ast_1, \bm{\theta}^\ast_2,\ldots,\bm{\theta}^\ast_{n_1}\}$ denote $n_1$ samples from the posterior distribution $p(\bm{\theta} \,|\, \bm{y})$ and $\{\tilde{\bm{\theta}}_1, \tilde{\bm{\theta}}_2,\ldots,\tilde{\bm{\theta}}_{n_2}\}$ denote $n_2$ samples from the proposal distribution $g(\bm{\theta})$. 

To use bridge sampling in practice, one has to specify the bridge function $h(\bm{\theta})$ and the proposal distribution $g(\bm{\theta})$.
For the bridge function $h(\bm{\theta})$, the \pkg{bridgesampling} package implements the optimal choice presented in \citet{MengWong1996} which minimizes the relative mean-squared error of the estimator.
Using this particular bridge function, the bridge sampling estimate of the marginal likelihood is obtained via an iterative scheme that updates an initial guess of the marginal likelihood $\hat{p}(\bm{y})^{(0)}$ until convergence \citep[for details, see][]{MengWong1996,GronauEtAltutorial}.
The estimate at iteration $t + 1$ is obtained as follows:
\begin{equation}
\label{eq:bridge_iterative_scheme}
\hat{p}(\bm{y})^{(t+1)} = \frac{\frac{1}{n_2}\sum\limits_{j = 1}^{n_2}\frac{l_{2, j}}{s_1 \thinspace l_{2, j} + s_2 \thinspace \hat{p}(\bm{y})^{(t)}}}{\frac{1}{n_1}\sum\limits_{i = 1}^{n_1}\frac{1}{s_1 \thinspace l_{1, i} + s_2 \thinspace \hat{p}(\bm{y})^{(t)}}},
\end{equation}
where $l_{1,i} = \frac{p(\bm{y} \mid \bm{\theta}^\ast_i) \, p(\bm{\theta}^\ast_i)}{g(\bm{\theta}^\ast_i)}$,  and $l_{2,j} = \frac{p(\bm{y} \mid \tilde{\bm{\theta}}_j) \, p(\tilde{\bm{\theta}}_j)}{g(\tilde{\bm{\theta}}_j)}$. In practice, a more numerically stable version of Equation~(\ref{eq:bridge_iterative_scheme}) is implemented that uses logarithms in combination with the \pkg{Brobdingnag} \proglang{R} package \citep{Brobdingnag} to avoid numerical under- and overflow (for details, see \citealp[Appendix B]{GronauEtAltutorial}).

The iterative scheme usually converges within a few iterations. Note that, crucially, $l_{1,i}$ and $l_{2,j}$ need only be computed once before the iterative updating scheme is started. In practice, evaluating $l_{1,i}$ and $l_{2,j}$ takes up most of the computational time. Luckily, $l_{1,i}$ and $l_{2,j}$ can be computed completely in parallel for each $i \in \left\{1,2,\ldots, n_1\right\}$ and each $j \in \left\{1,2,\ldots, n_2\right\}$, respectively. That is, in contrast to MCMC procedures, the evaluation of, for instance, $l_{1,i+1}$ does \emph{not} require one to evaluate $l_{1,i}$ first (since the posterior samples and proposal samples are already available). The \pkg{bridgesampling} package enables the user to compute $l_{1,i}$ and $l_{2,j}$ in parallel by setting the argument \code{cores} to an integer larger than one.
On Unix/macOS machines, this parallelization is implemented using the \pkg{parallel} package. On Windows machines this is achieved using the \pkg{snowfall} package \citep{snowfall}.\footnote{Due to technical limitations specific to Windows, this parallelization is not available for the \code{stanfit} and \code{stanreg} methods.}

After having specified the bridge function, one needs to choose the proposal distribution $g(\bm{\theta})$. The \pkg{bridgesampling} package implements two different choices: (a) a multivariate normal proposal distribution with mean vector and covariance matrix that match the respective posterior samples quantities and (b) a standard multivariate normal distribution combined with a \emph{warped} posterior distribution.\footnote{Note that other proposal distributions such as multivariate $t$ distributions are conceivable but are currently not implemented in the \pkg{bridgesampling} package.}
Both choices increase the efficiency of the estimator by making the proposal and the posterior distribution as similar as possible. Note that under the optimal bridge function, the bridge sampling estimator is robust to the relative tail behavior of the posterior and the proposal distribution. This stands in sharp contrast to the importance and the generalized harmonic mean estimator for which unwanted tail behavior produces estimators with very large or even infinite variances \citep[e.g.,][]{OwenZhou2000,FruehwirthSchnatter2004,GronauEtAltutorial}.

\subsection{Option I: the multivariate normal proposal distribution}
The first choice for the proposal distribution that is implemented in the \pkg{bridgesampling} package is a multivariate normal distribution with mean vector and covariance matrix that match the respective posterior samples quantities. This choice (henceforth ``the normal method'') generalizes to high dimensions and accounts for potential correlations in the joint posterior distribution. This proposal distribution is obtained by setting the argument \code{method = "normal"} in the \code{bridge_sampler} function; this is the default setting. This choice assumes that all parameters are allowed to range across the entire real line. In practice, this assumption may not be fulfilled for all components of the parameter vector, however, it is usually possible to transform the parameters so that this requirement is met.
This is achieved by transforming the original $p$-dimensional parameter vector $\bm{\theta}$ (which may contain components that range only across a subset of $\mathbb{R}$) to a new parameter vector $\bm{\xi}$ (where all components are allowed to range across the entire real line) using a diffeomorphic vector-valued function $f$ so that $\bm{\xi} = f(\bm{\theta})$. By the change-of-variable rule, the posterior density with respect to the new parameter vector $\bm{\xi}$ is given by:
\begin{equation}
\label{eq:transformed_posterior}
p(\bm{\xi} \mid \bm{y}) = p_{\bm{\theta}}(f^{-1}(\bm{\xi}) \mid \bm{y}) \, \left\lvert\det\left[J_{f^{-1}}(\bm{\xi})\right]\right\rvert,
\end{equation}
where $p_{\bm{\theta}}(f^{-1}(\bm{\xi}) \mid \bm{y})$ refers to the untransformed posterior density with respect to $\bm{\theta}$ evaluated for $f^{-1}(\bm{\xi}) = \bm{\theta}$. $J_{f^{-1}}(\bm{\xi})$ denotes the Jacobian matrix with the element in the $i$-th row and $j$-th column given by $\frac{\partial \theta_i}{\partial \xi_j}$. Crucially, the posterior density with respect to $\bm{\xi}$ retains the normalizing constant of the posterior density with respect to $\bm{\theta}$; hence, one can select a convenient transformation without changing the normalizing constant. Note that in order to apply a transformation no new samples are required; instead the original samples can simply be transformed using the function $f$.

In principle, users can select transformations themselves. Nevertheless, the \pkg{bridgesampling} package comes with a set of built-in transformations (see Table~\ref{tab:transformations}), allowing the user to work with the model in a familiar parameterization. When the user then supplies a named vector with lower and upper bounds for the parameters (arguments \code{lb} and \code{ub}, respectively), the package internally transforms the relevant parameters and adjusts the expressions by the Jacobian term. Furthermore, as will be elaborated upon below, when the model is fitted in \proglang{Stan}, the \pkg{bridgesampling} package takes advantage of the rich class of \proglang{Stan} transformations.

The transformations built into the \pkg{bridgesampling} package are useful whenever each component of the parameter vector can be transformed separately.\footnote{Thanks to a recent pull request by Kees Mulder, the \pkg{bridgesampling} package now also supports a more complicated case in which multiple parameters are constrained jointly (i.e., simplex parameters). This pull request also added support for circular parameters.} In this scenario, there are four possible cases per parameter: (a) the parameter is unbounded; (b) the parameter has a lower bound (e.g., variance parameters); (c) the parameter has an upper bound; and (d) the parameter has a lower and an upper bound (e.g., probability parameters). As shown in Table~\ref{tab:transformations}, in case (a) the identity (i.e., no) transformation is applied. In case (b) and (c), logarithmic transformations are applied to transform the parameter to the real line. In case (d) a probit transformation is applied. Note that internally, the posterior density is automatically adjusted by the relevant Jacobian term. Since each component is transformed separately, the resulting Jacobian matrix will be diagonal. This is convenient since it implies that the absolute value of the determinant is the product of the absolute values of the diagonal entries of the Jacobian matrix:
\begin{equation}
\label{eq:det_diag_Jacobian}
\left\lvert\det\left[J_{f^{-1}}(\bm{\xi})\right]\right\rvert = \prod_{i = 1}^{p} \left\lvert\frac{\partial \theta_i}{\partial \xi_i}\right\rvert.
\end{equation}

\begin{table}[bt]
	\centering
		\normalsize
		\caption{Overview of built-in transformations in the \pkg{bridgesampling} package. $l$ denotes a parameter lower bound and $u$ denotes an upper bound. $\Phi(\cdot)$ denotes the cumulative distribution function (cdf) and $\phi(\cdot)$ the probability density function (pdf) of the normal distribution.}
	\begin{tabular}{llll}
		\toprule
		\textbf{Type} & \textbf{Transformation} & \textbf{Inv.-Transformation} & \textbf{Jacobian Contribution} \\
        \hline
        unbounded & $\xi_i = \theta_i$ & $\theta_i = \xi_i$ & $\left\lvert\frac{\partial \theta_i}{\partial \xi_i}\right\rvert = 1$ \\ [1em]
        lower-bounded & $\xi_i = \log\left(\theta_i - l\right)$ & $\theta_i = \exp\left(\xi_i\right) + l$ & $\left\lvert\frac{\partial \theta_i}{\partial \xi_i}\right\rvert = \exp\left(\xi_i\right)$ \\ [1em]
        upper-bounded & $\xi_i = \log\left(u - \theta_i\right)$ & $\theta_i = u - \exp\left(\xi_i\right)$ & $\left\lvert\frac{\partial \theta_i}{\partial \xi_i}\right\rvert = \exp\left(\xi_i\right)$ \\ [1em]
        double-bounded & $\xi_i = \Phi^{-1}\left(\frac{\theta_i - l}{u - l}\right)$ & $\theta_i = \left(u - l\right)\Phi\left(\xi_i\right) + l$ & $\left\lvert\frac{\partial \theta_i}{\partial \xi_i}\right\rvert = \left(u - l\right) \phi\left(\xi_i\right)$\\
		\bottomrule
        \label{tab:transformations}
	\end{tabular}
\end{table}

Once all posterior samples have been transformed to the real line, a multivariate normal distribution is fitted using method-of-moments. On a side note, bridge sampling may underestimate the marginal likelihood when the same posterior samples are used both for fitting the proposal distribution and for the iterative updating scheme (i.e., Equation~(\ref{eq:bridge_iterative_scheme})). Hence, as recommended by \citet{OverstallForster2010}, the \pkg{bridgesampling} package divides each MCMC chain into two halves, using the first half for fitting the proposal distribution and the second half for the iterative updating scheme.

\subsection{Option II: warping the posterior distribution}
The second choice for the proposal distribution that is implemented in the \pkg{bridgesampling} package is a standard multivariate normal distribution in combination with a \emph{warped} posterior distribution. The goal is still to match the posterior and the proposal distribution as closely as possible. However, instead of manipulating the proposal distribution, it is fixed to a standard multivariate normal distribution, and the posterior distribution is manipulated (i.e., warped). Crucially, the warped posterior density retains the normalizing constant of the original posterior density. The general methodology is referred to as Warp bridge sampling \citep{MengSchilling2002}.

There exist several variants of Warp bridge sampling; in the \pkg{bridgesampling} package, we implemented Warp-III bridge sampling \citep{MengSchilling2002,Overstall2010thesis,GronauEtAlwarp3} which can be used by setting \code{method = "warp3"}. This version matches the first three moments of the posterior and the proposal distribution. That is, in contrast to the simpler normal method described above, Warp-III not only matches the mean vector and the covariance matrix of the two distributions, but also the skewness. Consequently, when the posterior distribution is skewed, Warp-III may result in an estimator that is less variable. When the posterior distribution is symmetric, both Warp-III and the normal method should yield estimators that are about equally efficient. Hence, in principle, Warp-III should always provide estimates that are at least as precise as the normal method. However, the Warp-III method also takes about twice as much time to execute as the normal method; the reason for this is that Warp-III sampling results in a mixture density \citep[for details, see][]{Overstall2010thesis,GronauEtAlwarp3} which requires that the unnormalized posterior density is evaluated twice as often as in the normal method.

\begin{figure}[!tb]
	\begin{center}
		\includegraphics{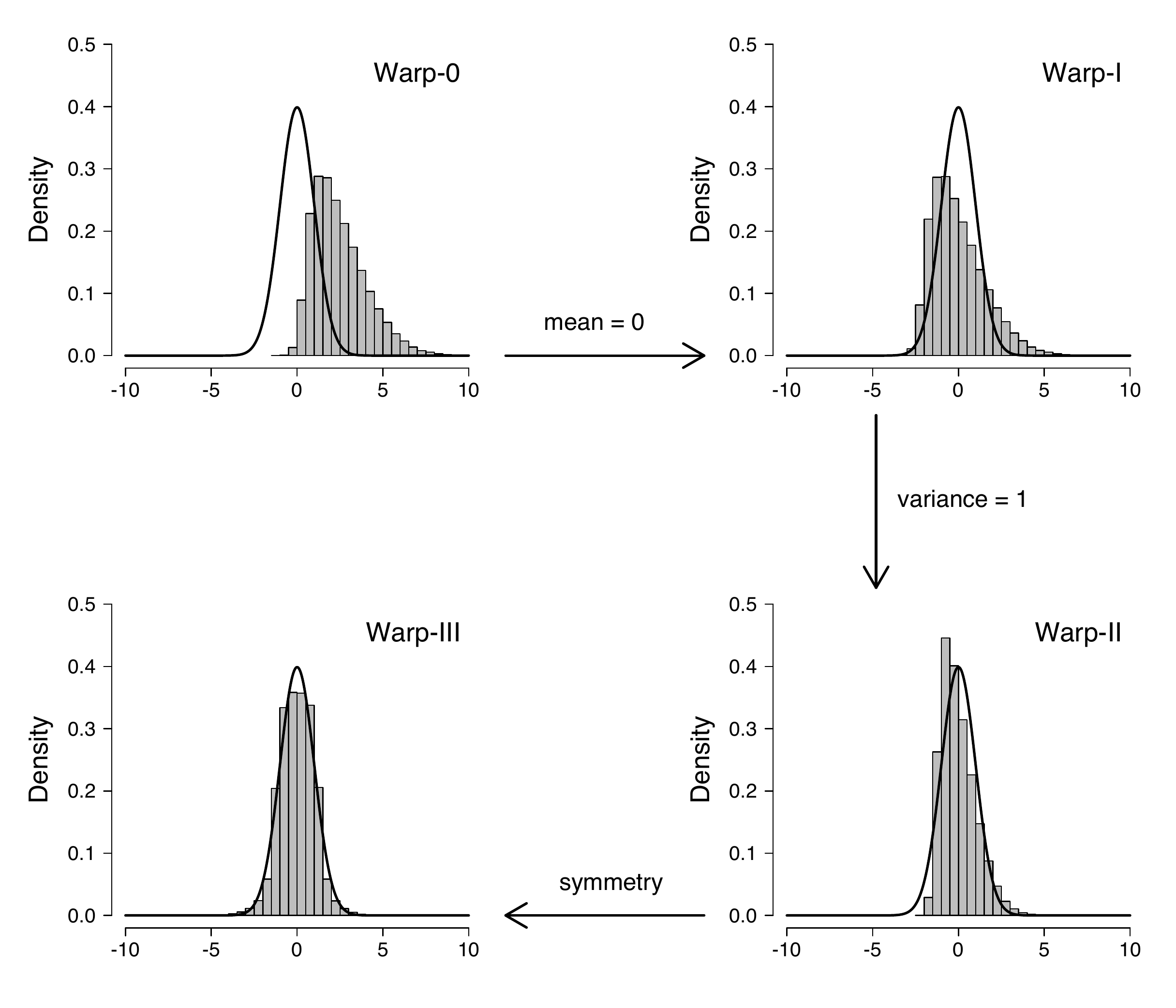}
		\caption{Illustration of the warping procedure. The black solid line shows the standard normal proposal distribution and the gray histogram shows the posterior samples. Available at \url{https://tinyurl.com/y7owvsz3} under CC license \url{https://creativecommons.org/licenses/by/2.0/} (see also \citealp{GronauEtAlwarp3,GronauEtAlLBA}).}
		\label{fig:warping}
	\end{center}
\end{figure}

Figure~\ref{fig:warping} illustrates the intuition for the warping procedure in the univariate case. The gray histogram in the top-left panel depicts skewed posterior samples, the solid black line the standard normal proposal distribution. The Warp-III procedure effectively standardizes the posterior samples so that they have mean zero (top-right panel) and variance one (bottom-right panel), and then attaches a minus sign with probability 0.5 to the samples which achieves symmetry (bottom-left panel). This intuition naturally generalizes to the multivariate case.
Starting with posterior samples that can range across the entire real line (i.e., $\bm{\xi}$) the multivariate Warp-III procedure is based on the following stochastic transformation:
\begin{equation}
\label{eq:warp3_stochastic_transfromation}
\bm{\eta} = \underbrace{b}_{\textstyle\text{symmetry}} \times \underbrace{\bm{R}^{-1}}_{\textstyle\text{covariance $\bm{I}$}} \times \;\; \underbrace{(\bm{\xi} - \bm{\mu})}_{\textstyle \text{mean $\bm{0}$}},
\end{equation}
where $b \sim \mathcal{B}(0.5)$ on $\{-1, 1\}$ and $\bm{\mu}$ corresponds to the expected value of $\bm{\xi}$ (i.e., the mean vector).\footnote{$\mathcal{B}(\theta)$ denotes a Bernoulli distribution with success probability $\theta$.} The matrix $\bm{R}$ is obtained via the Cholesky decomposition of the covariance matrix of $\bm{\xi}$, denoted as $\bm{\Sigma}$, hence, $\bm{\Sigma} = \bm{R} \bm{R}^\top$. Bridge sampling is then applied using this warped posterior distribution in combination with a standard multivariate normal distribution.

\subsection{Estimation error}
Once the marginal likelihood has been estimated, the user can obtain an estimate of the estimation error in a number of different ways. One method is to use the \code{error_measures} function which is an \code{S3} generic. Note that the \code{summary} method for objects returned by \code{bridge\_sampler} internally calls the \code{error_measures} function and thus provides a convenient summary of the estimated log marginal likelihood and the estimation uncertainty.
For marginal likelihoods estimated with the \code{"normal"} method and \code{repetitions = 1}, the \code{error\_measures} function provides an approximate relative mean-squared error of the marginal likelihood estimate, an approximate coefficient of variation, and an approximate percentage error. The relative mean-squared error of the marginal likelihood estimate is given by:
\begin{equation}
\label{eq:RE2}
\text{RE}^2 = \frac{ \mathbb{E}\left [\big(\hat p(\bm{y}) - p(\bm{y})\big)^2 \right ]}{p(\bm{y})^2}.
\end{equation}
The \pkg{bridgesampling} package computes an approximate relative mean-squared error of the marginal likelihood estimate based on the derivation by \citet{FruehwirthSchnatter2004} which takes into account that the samples from the proposal distribution are independent, whereas the samples from the posterior distribution may be autocorrelated (e.g., when using MCMC sampling procedures).

Under the assumption that the bridge sampling estimator $\hat p(\bm{y})$ is unbiased, the square root of the expected relative mean-squared error (Equation~(\ref{eq:RE2})) can be interpreted as the coefficient of variation (i.e., the ratio of the standard deviation and the mean). To facilitate interpretation, the \pkg{bridgesampling} package also provides a percentage error which is obtained by simply converting the coefficient of variation to a percentage.

Note that the \code{error\_measures} function can currently not be used to obtain approximate errors for the \code{"warp3"} method with \code{repetitions = 1}. The reason is that, in our experience, the approximate errors appear to be unreliable in this case.

There are two further methods for assessing the uncertainty of the marginal likelihood estimate. These methods are computationally more costly than computing approximate errors, but are available for both the \code{"normal"} method and the \code{"warp3"} method. The first option is to set the \code{repetitions} argument of the \code{bridge_sampler} function to an integer larger than one. This allows the user to obtain an empirical estimate of the variability across repeated applications of the method. Applying the \code{error\_measures} function to the output of the \code{bridge\_sampler} function that has been obtained with \code{repetitions} set to an integer large than one provides the user with the minimum/maximum log marginal likelihood estimate across repetitions and the interquartile range of the log marginal likelihood estimates. Note that this procedure assesses the uncertainty of the estimate conditional on the posterior samples, that is, in each repetition new samples are drawn from the proposal distribution, but the posterior samples are fixed across repetitions. 

In case the user is able to easily draw new samples from the posterior distribution, the second option is to repeatedly call the \code{bridge_sampler} function, each time with new posterior samples. This way, the user obtains an empirical assessment of the variability of the estimate which takes into account both uncertainty with respect to the samples from the proposal and also from the posterior distribution. If computationally feasible, we recommend this method for assessing the estimation error of the marginal likelihood.

After having outlined the underlying bridge sampling algorithm, we next demonstrate the capabilities of the \pkg{bridgesampling} package using three examples. Additional examples are available as vignettes at: \url{https://cran.r-project.org/package=bridgesampling}

\section{Toy example: Bayesian $t$-test}

\begin{figure}[!tb]
 \begin{center}
	\includegraphics{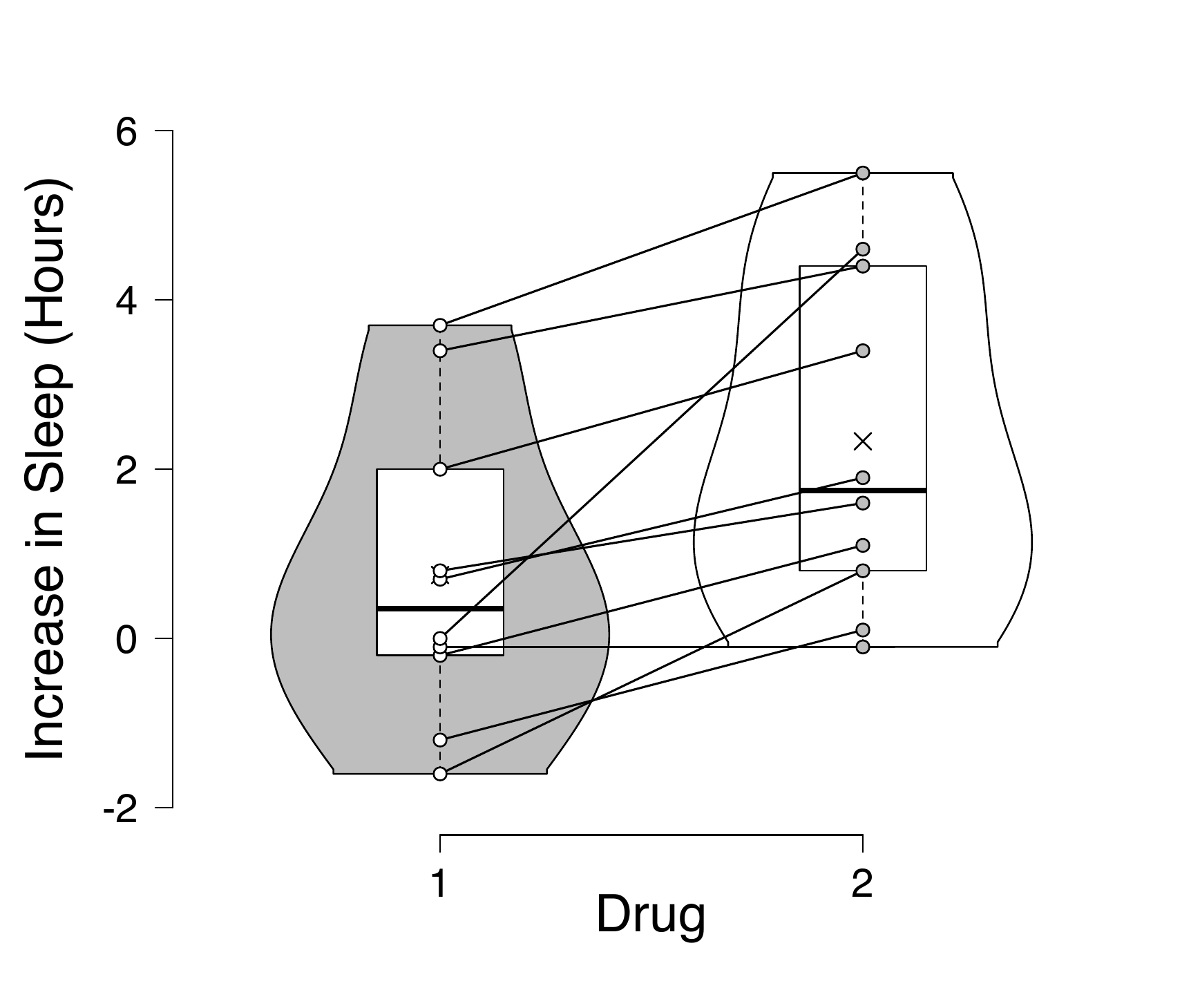}
	\caption{The sleep data set \citep{CushnyPeebles1905}. The left violin plot displays the distribution of the increase in sleep (in hours) of the ten patients for the first drug, the right violin plot displays the distribution of the increase in sleep (in hours) of the ten patients for the second drug. Boxplots and the individual observations are superimposed. Observations for the same participant are connected by a line. Figure available at \url{https://tinyurl.com/yalskr23} under CC license \url{https://creativecommons.org/licenses/by/2.0/}.}
	\label{fig:sleep}
	 \end{center}
\end{figure}

We start with a simple statistical example: a Bayesian paired-samples $t$-test \citep{Jeffreys1961,RouderEtAl2009Ttest,LyEtAl2016,GronauEtAlarxiv}. We use \proglang{R}'s 
\code{sleep} data set \citep{CushnyPeebles1905} which contains measurements for the effect of two soporific drugs on ten patients. Two different drugs where administered to the same ten patients and the dependent measure was the average number of hours of sleep gained compared to a control night in which no drug was administered. Figure~\ref{fig:sleep} shows the increase in sleep (in hours) of the ten patients for each of the two drugs.
To test whether the two drugs differ in effectiveness, we can conduct a Bayesian paired-samples $t$-test.

The null hypothesis $\mathcal{H}_0$ states that the $n$ difference scores $d_i, \, i = 1, 2, \ldots, n$, where $n = 10$, follow a normal distribution with mean zero and variance $\sigma^2$, that is, $d_i \sim \mathcal{N}(0, \sigma^2)$.
The alternative hypothesis $\mathcal{H}_1$ states that the difference scores follow a normal distribution with mean $\mu = \sigma \delta$, where $\delta$ denotes the standardized effect size, and variance $\sigma^2$, that is, $d_i \sim \mathcal{N}(\sigma \delta, \sigma^2)$. Jeffreys's prior is assigned to the variance $\sigma^2$ so that $p(\sigma^2) \propto 1/\sigma^2$ and a zero-centered Cauchy prior with scale parameter $r = 1/\sqrt{2}$ is assigned to the standardized effect size $\delta$ \citep[for details, see][]{RouderEtAl2009Ttest, LyEtAl2016, MoreyRouderBayesFactorPackage}.

In this example, we are interested in computing the Bayes factor  $\text{BF}_{10}$ which quantifies how much more likely the data are under $\mathcal{H}_1$ (i.e., there is a difference between the two drugs) than under $\mathcal{H}_0$ (i.e., there is no difference between the two drugs) by using the \pkg{bridgesampling} package. For this example, the Bayes factor can also be easily computed using the \pkg{BayesFactor} package \citep{MoreyRouderBayesFactorPackage}, allowing us to compare the results from the \pkg{bridgesampling} package to the correct answer.

The first step is to obtain posterior samples. In this example, we use \proglang{JAGS} in order to sample from the models.
Here we focus on how to compute the log marginal likelihood for $\mathcal{H}_1$. The steps for obtaining the log marginal likelihood for $\mathcal{H}_0$ are analogous.
After having specified the model corresponding to $\mathcal{H}_1$ as the character string \code{code\_H1}, posterior samples can be obtained using the \pkg{R2jags} package \citep{R2jags} as follows:\footnote{The complete code (including the \proglang{JAGS} models and the code for $\mathcal{H}_0$) can be found in the supplemental material and also on the Open Science Framework: \url{https://osf.io/3yc8q/}.}
\begin{Sinput}
R> library("R2jags")
R> data("sleep")
R> y <- sleep$extra[sleep$group == 1]
R> x <- sleep$extra[sleep$group == 2]
R> d <- x - y  # compute difference scores
R> n <- length(d)
R> set.seed(1)  
R> jags_H1 <- jags(data = list(d = d, n = n, r = 1 / sqrt(2)),
+                  parameters.to.save = c("delta", "inv_sigma2"),
+                  model.file = textConnection(code_H1), n.chains = 3,
+                  n.iter = 16000, n.burnin = 1000, n.thin = 1)
\end{Sinput}
Note the relatively large number of posterior samples; reliable estimates for the quantities of interest in testing usually necessitate many more posterior samples than are required for estimation. As a rule of thumb, we suggest that testing requires about an order of magnitude more posterior samples than estimation.

Next, we need to specify a function that take as input a named vector with parameter values and a data object, and returns the log of the unnormalized posterior density (i.e., the log of the integrand in Equation~(\ref{eq:marginal_likelihood})). This function is easily specified by inspecting the \proglang{JAGS} model. As a heuristic, one only needs to consider the model code where a ``$\sim$'' sign appears.
The log of the densities on the right-hand side of these ``$\sim$'' symbols needs to be evaluated for the relevant quantities and then these log density values are summed.\footnote{This heuristic assumes that the model does not include other random quantities that are generated during sampling, such as posterior predictives.}
Using this heuristic, we obtain the following unnormalized log posterior density function for $\mathcal{H}_1$:
\begin{Sinput}
R> log_posterior_H1 <- function(pars, data) {
+   delta <- pars["delta"]            # extract parameter
+   inv_sigma2 <- pars["inv_sigma2"]  # extract parameter
+   sigma <- 1 / sqrt(inv_sigma2)     # convert precision to sigma
+   out <-
+     dcauchy(delta, scale = data$r, log = TRUE) +         # prior
+     dgamma(inv_sigma2, 0.0001, 0.0001, log = TRUE) +     # prior
+     sum(dnorm(data$d, sigma * delta, sigma, log = TRUE)) # likelihood
+   return(out)
+ }
\end{Sinput}

The final step before we can compute the log marginal likelihoods is to specify named vectors with the parameter bounds:
\begin{Sinput}
R> lb_H1 <- rep(-Inf, 2)
R> ub_H1 <- rep(Inf, 2)
R> names(lb_H1) <- names(ub_H1) <- c("delta", "inv_sigma2")
R> lb_H1[["inv_sigma2"]] <- 0
\end{Sinput}

The log marginal likelihood for $\mathcal{H}_1$ can then be obtained by calling the \code{bridge_sampler} function as follows:
\begin{Sinput}
R> library("bridgesampling")
R> set.seed(12345)
R> bridge_H1 <- bridge_sampler(samples = jags_H1,
+                              log_posterior = log_posterior_H1,
+                              data = list(d = d, n = n, r = 1 / sqrt(2)),
+                              lb = lb_H1, ub = ub_H1)
\end{Sinput}
We obtain:
\begin{Sinput}
R> print(bridge_H1)
\end{Sinput}
\begin{Soutput}
Bridge sampling estimate of the log marginal likelihood: -27.17103
Estimate obtained in 5 iteration(s) via method "normal".
\end{Soutput}
Note that by default, the \code{"normal"} bridge sampling method is used.

Next, we can use the \code{error_measures} function to obtain an approximate percentage error of the estimate:
\begin{Sinput}
R> error_measures(bridge_H1)$percentage
\end{Sinput}
\begin{Soutput}
[1] "0.087%"
\end{Soutput}
The small approximate percentage error indicates that the marginal likelihood has been estimated reliably.
As mentioned before, we can use the \code{summary} method to obtain a convenient summary of the bridge sampling estimate and the estimation error. We obtain:
\begin{Sinput}
R> summary(bridge_H1)
\end{Sinput}
\begin{Soutput}
Bridge sampling log marginal likelihood estimate 
(method = "normal", repetitions = 1):

 -27.17103

Error Measures:

 Relative Mean-Squared Error: 7.564225e-07
 Coefficient of Variation: 0.0008697255
 Percentage Error: 0.087%

Note:
All error measures are approximate.
\end{Soutput}
After having computed the log marginal likelihood estimate for $\mathcal{H}_0$ in a similar fashion, we can compute the Bayes factor for $\mathcal{H}_1$ over $\mathcal{H}_0$ using the \code{bf} function:
\begin{Sinput}
R> bf(bridge_H1, bridge_H0)
\end{Sinput}
\begin{Soutput}
Estimated Bayes factor in favor of bridge_H1 over bridge_H0: 17.26001
\end{Soutput}
Hence, the observed data are about 17 times more likely under $\mathcal{H}_1$ (which assigns the standardized effect size $\delta$ a zero-centered Cauchy prior with scale $r = 1/\sqrt{2}$) than under $\mathcal{H}_0$ (which fixes $\delta$ to zero). This is strong evidence for a difference in effectiveness between the two drugs \citep[Appendix I]{Jeffreys1939}.
The estimated Bayes factor closely matches the Bayes factor obtained with the \pkg{BayesFactor} package (i.e., $\text{BF}_{10} = 17.259$).

\section[A ``black box'' Stan interface]{A ``black box'' \proglang{Stan} interface}

The previous section demonstrated how the \pkg{bridgesampling} package can be used to estimate the marginal likelihood for models coded in \proglang{JAGS}.
For custom samplers, the steps needed to compute the marginal likelihood are the same. What is required is (a) an object with posterior samples; (b) a function that computes the log of the unnormalized posterior density; (c) the data; and (d) parameter bounds.
A crucial step is the specification of the unnormalized log posterior density function.
For applied researchers, this step may be challenging and error-prone, whereas for experienced statisticians it might be tedious and cumbersome, especially for complex models with a hierarchical structure.

In order to facilitate the computation of the marginal likelihood even further, the \pkg{bridgesampling} package contains an interface to the generic sampling software \proglang{Stan} \citep{CarpenterEtAl2017}. Assisted by the \pkg{rstan} package \citep{rstan}, this interface allows users to skip steps (b)-(d) above. Specifically, users who fit their models in \proglang{Stan} (in a way that retains the constants, as is detailed below) can obtain an estimate of the marginal likelihood by simply passing the \code{stanfit} object to the \code{bridge_sampler} function.

The implementation of this ``black box'' functionality profited from the fact that, just as the \pkg{bridgesampling} package, \proglang{Stan}'s No-U-Turn sampler internally operates on unconstrained parameters \citep{HoffmanGelman2014,stanmanual}. The \pkg{rstan} package provides access to these unconstrained parameters and the corresponding log of the unnormalized posterior density. This means that users can fit models with parameter types that have more complicated constraints than those currently built into \pkg{bridgesampling} (e.g., covariance/correlation matrices) without having to hand-code the appropriate transformations.

As mentioned above, in order to use the \pkg{bridgesampling} package in combination with \proglang{Stan} the models need to be implemented in a way that retains the constants. This can be achieved relatively easily: instead of writing, for instance, \code{y ~ normal(mu, sigma)} or \code{y ~ bernoulli(theta)}, one needs to write
\begin{Sinput}
target += normal_lpdf(y | mu, sigma);
\end{Sinput}
and 
\begin{Sinput}
target += bernoulli_lpmf(y | theta);
\end{Sinput}
That is, one starts with the fixed expression \code{target +=} which is then followed by the name of the distribution (e.g., \code{normal}). The name of the distribution is followed by \code{_lpdf} for continuous distributions and \code{_lpmf} for discrete distributions. Finally, in parentheses, there is the variable that was to the left of the ``\code{~}'' sign (here, \code{y}), then a ``\code{|}'' sign, and finally the arguments of the distribution. This achieves that the user specifies the log target density (in this case, the log of the unnormalized posterior density) in a way that retains the constants of the involved distributions.

Note that in case the distributions are truncated, the user needs to code the correct renormalization. For instance,  a normal distribution with upper truncation at \code{upper} is implemented as follows 
\begin{Sinput}
target += normal_lpdf(y | mu, sigma) - normal_lcdf(upper | mu, sigma);
\end{Sinput}
where the function \code{normal_lcdf} yields the log of the cumulative distribution function (cdf) of the normal distribution.
Likewise, a normal distribution with lower truncation at \code{lower} is obtained as
\begin{Sinput}
target += normal_lpdf(y | mu, sigma) - normal_lccdf(lower | mu, sigma);
\end{Sinput}
where \code{normal\_lccdf} yields the log of the complementary cumulative distribution function (ccdf) of the normal distribution (i.e., the log of one minus the cumulative distribution function of the normal distribution).
A normal distribution with lower truncation point \code{lower} and upper truncation point \code{upper} can be implemented as follows:
\begin{Sinput}
target += normal_lpdf(y | mu, sigma) -
          log_diff_exp(normal_lcdf(upper | mu, sigma),
                       normal_lcdf(lower | mu, sigma));
\end{Sinput}
where \code{log_diff_exp(a, b)} is a numerically more stable version of the operation
$\log\left(\exp\left(a\right) - \exp\left(b\right)\right)$.
Note that when implementing a truncated distribution, it is of course also important to give the variable of interest the correct bounds. For instance, for the last example where \code{y} has a lower truncation at \code{lower} and an upper truncation at \code{upper} the variable \code{y} should be declared as\footnote{Note that we assumed that \code{y} is a scalar. In general, \code{y} could also be declared as a vector or an array in \proglang{Stan}. In this case, the term that is subtracted for renormalization would need to be multiplied by the number of elements of \code{y}. For example, for the case of an upper truncation and a vector \code{y} of length \code{k} the code would need to be changed to:
\code{target += normal\_lpdf(y | mu, sigma) - k * normal\_lcdf(upper | mu, sigma);} For another example, see the code for ``\proglang{Stan} example 2''.}
\begin{Sinput}
real<lower = lower, upper = upper> y;
\end{Sinput}
For more details about how to implement truncated distributions in \proglang{Stan} we refer the user to the \proglang{Stan} manual \citep[section 5.3, ``Truncated Distributions'']{stanmanual}. 

In sum, the \pkg{bridgesampling} package enables users to obtain an estimate of the marginal likelihood for any \proglang{Stan} model (programmed to retain the constants) simply by passing the \code{stanfit} object to the \code{bridge_sampler} function. Next we demonstrate this functionality using two prototypical examples in Bayesian model selection.

\subsection[Stan example 1: Bayesian GLMM]{\proglang{Stan} example 1: Bayesian GLMM}

\begin{figure}[!tb]
 \begin{center}
	\includegraphics{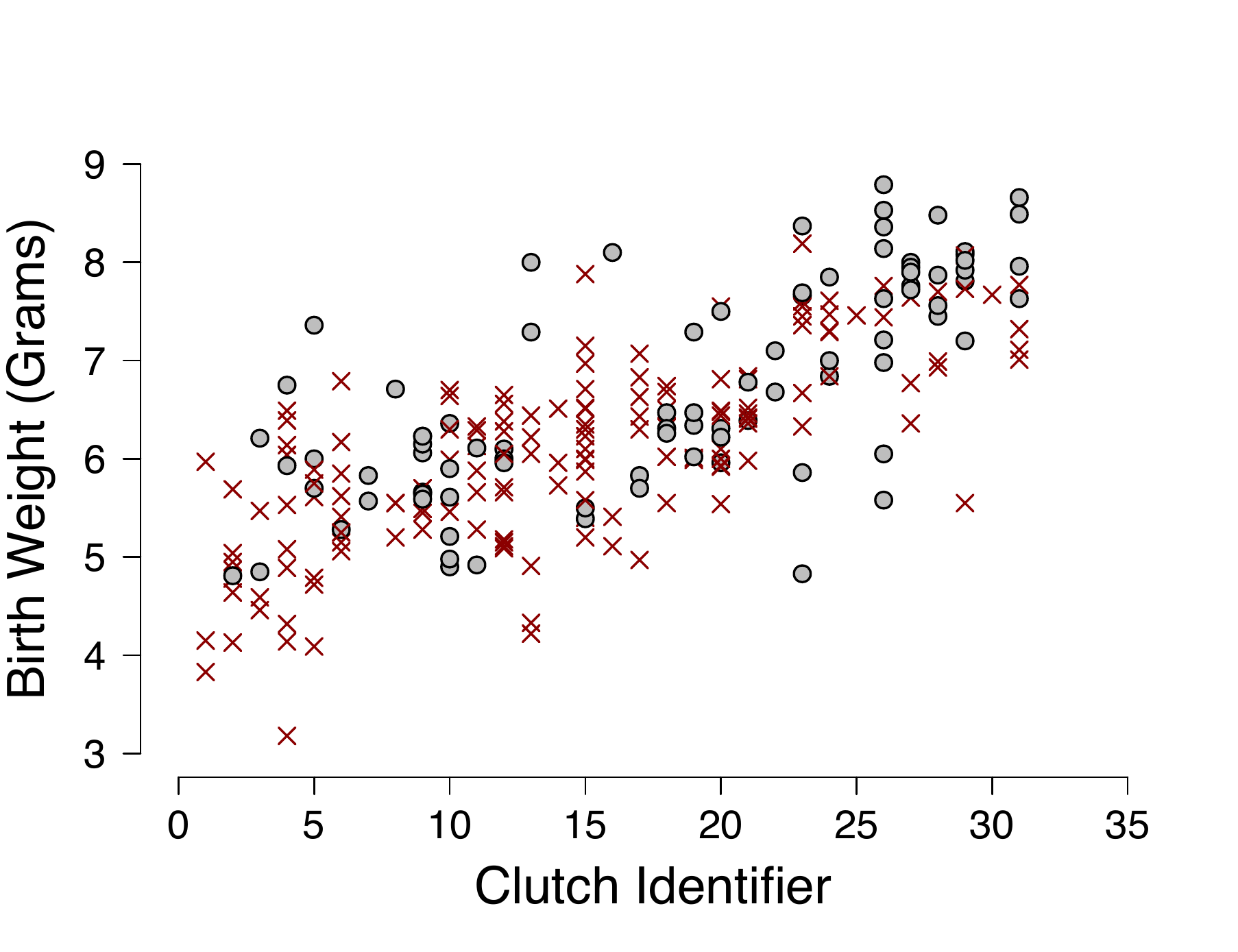}
	\caption{Data for 244 newborn turtles \citep{JanzenEtAl2000}. Birth weight is plotted against clutch membership. The clutches have been ordered according to their mean birth weight. Dots indicate turtles who survived and red crosses indicate turtles who died. Figure inspired by \citet{SinharayStern2005}. Figure available at \url{https://tinyurl.com/yagfxrbw} under CC license \url{https://creativecommons.org/licenses/by/2.0/}.}
	\label{fig:turtles}
	 \end{center}
\end{figure}

The first example features a generalized linear mixed model (GLMM) applied to the turtles data set \citep{JanzenEtAl2000}.\footnote{Data were obtained from \citet{OverstallForster2010} and made available in the \pkg{bridgesampling} package with permission from the original authors.} This data set is included in the \pkg{bridgesampling} package and contains information about 244 newborn turtles from 31 different clutches. For each turtle, the data set includes information about survival status (0 = died, 1 = survived), birth weight in grams, and clutch (family) membership (indicated by a number between one and 31). 
Figure~\ref{fig:turtles} displays a scatterplot of clutch membership and birth weight. The clutches have been ordered according to mean birth weight.
Dots indicate turtles who survived and red crosses indicate turtles who died.
This data set has been analyzed in the context of Bayesian model selection before, allowing us to compare the results from the \pkg{bridgesampling} package to the results reported in the literature \citep[e.g.,][]{SinharayStern2005,OverstallForster2010}.

Here we focus on the model comparison that was conducted in \citet{SinharayStern2005}. The data set was analyzed using a probit regression model of the form:
\begin{equation}
\begin{split}
y_i &\sim  \mathcal{B}(\Phi(\alpha_0 + \alpha_1 x_{i} + b_{\text{clutch}_i})), \hspace{1.68cm} i = 1,2,\ldots,N\\
b_j &\sim \mathcal{N}(0, \sigma^2), \hspace{11.8em} j = 1,2,\ldots, C,
\end{split}
\end{equation}
where $y_i$ denotes the survival status of the $i$-th turtle (i.e., 0 = died, 1 = survived), $x_i$ denotes the birth weight (in grams) of the $i$-th turtle, $\text{clutch}_i \in \{1,2,\ldots, C\}, \, i = 1,2,\ldots, N$, indicates the clutch to which the $i$-th turtle belongs, $C$ denotes the number of clutches, and $b_{\text{clutch}_i}$ denotes the random effect for the clutch to which the $i$-th turtle belongs. Furthermore, $\Phi(\cdot)$ denotes the cumulative distribution function (cdf) of the normal distribution.  \citet{SinharayStern2005} investigated the question whether there is an effect of clutch membership, that is, they tested the null hypothesis $\mathcal{H}_0: \sigma^2 = 0$. The following priors where assigned to the model parameters:
\begin{equation}
\begin{split}
\alpha_0 &\sim \mathcal{N}(0, 10),\\
\alpha_1 &\sim \mathcal{N}(0, 10),\\
p(\sigma^2) &= \left(1 + \sigma^2\right)^{-2}.
\end{split}
\end{equation}
\citet{SinharayStern2005} computed the Bayes factor in favor of the null hypothesis $\mathcal{H}_0: \sigma^2 = 0$ versus the alternative hypothesis $\mathcal{H}_1: p(\sigma^2) = \left(1 + \sigma^2\right)^{-2}$ using different methods and they reported a ``true'' Bayes factor of $\text{BF}_{01} = 1.273$ (based on extensive numerical integration). Here we examine the extent to which we can reproduce the Bayes factor using the \pkg{bridgesampling} package.

After having implemented the \proglang{Stan} models as character strings \code{H0\_code} and \code{H1\_code}, the next step is to run \proglang{Stan} and obtain the posterior samples:\footnote{The complete code can be found in the supplemental material, on the Open Science Framework (\url{https://osf.io/3yc8q/}), and is also available at \code{?turtles}. Note that the results are dependent on the compiler and the optimization settings. Thus, even with identical seeds results can differ slightly from the ones reported here.}
\begin{Sinput}
R> library("bridgesampling")
R> library("rstan")
R> data("turtles")
R> set.seed(1)
R> stanfit_H0 <- stan(model_code = H0_code,
+                     data = list(y = turtles$y,
+                     x = turtles$x, N = nrow(turtles)),
+                     iter = 15500, warmup = 500,
+                     chains = 4, seed = 1)
R> stanfit_H1 <- stan(model_code = H1_code,
+                     data = list(y = turtles$y,
+                     x = turtles$x, N = nrow(turtles),
+                     C = max(turtles$clutch),
+                     clutch = turtles$clutch),
+                     iter = 15500, warmup = 500,
+                     chains = 4, seed = 1)
\end{Sinput}

With these \proglang{Stan} objects in hand, estimates of the log marginal likelihoods are obtained by simply passing the objects to the \code{bridge_sampler} function:
\begin{Sinput}
R> set.seed(1)
R> bridge_H0 <- bridge_sampler(stanfit_H0)
R> bridge_H1 <- bridge_sampler(stanfit_H1)
\end{Sinput}

The Bayes factor in favor of $\mathcal{H}_0$ over $\mathcal{H}_1$ can then be obtained as follows:
\begin{Sinput}
R> bf(bridge_H0, bridge_H1)
\end{Sinput}
\begin{Soutput}
Estimated Bayes factor in favor of bridge_H0 over bridge_H1: 1.27151
\end{Soutput}
This value is close to that of 1.273 reported in \citet{SinharayStern2005}. The data are only slightly more likely under $\mathcal{H}_0$ than under $\mathcal{H}_1$, suggesting that the data do not warrant strong claims about whether or not clutch membership affects survival.

The precision of the estimates for the marginal likelihoods can be obtained as follows:
\begin{Sinput}
R> error_measures(bridge_H0)$percentage
\end{Sinput}
\begin{Soutput}
 [1] "0.00972%"
\end{Soutput}
\begin{Sinput}
R> error_measures(bridge_H1)$percentage
\end{Sinput}
\begin{Soutput}
[1] "0.348%"
\end{Soutput}
These error percentages indicate that both marginal likelihoods have been estimated accurately, but -- as expected -- the marginal likelihood for the more complicated model with random effects (i.e., $\mathcal{H}_1$) has the larger estimation error.

\subsection[Stan example 2: Bayesian factor analysis]{\proglang{Stan} example 2: Bayesian factor analysis}

The second example concerns Bayesian factor analysis. In particular, we determine the number of relevant latent factors by implementing the Bayesian factor analysis model proposed by \cite{LopesWest2004}. 
The model assumes that there are $t,\; t = 1,2,\ldots,T$, observations on each of $m$ variables. That is, each observation $\bm{y}_t$ is an $m$-dimensional vector.
The $k$-factor model -- where $k$ denotes the number of factors -- relates each of the $T$ observations $\bm{y}_t$ to a latent $k$-dimensional vector $\bm{f}_t$ which contains for observation $t$ the values on the latent factors, as follows:\footnote{Note that the model assumes that the observations are zero-centered.}
\begin{equation}
\begin{split}
\bm{y}_t \mid \bm{f}_t &\sim \mathcal{N}_m\left(\bm{\beta}\bm{f}_t, \bm{\Sigma}\right)\\
\bm{f}_t &\sim \mathcal{N}_k\left(\bm{0}_k, \bm{I}_k\right),
\end{split}
\end{equation}
where $\bm{\beta}$ denotes the $m \times k$ factor loadings matrix\footnote{We use the original notation by \citet{LopesWest2004} who denoted the factor loadings matrix with a lower-case letter. In the remainder of the article, matrices are denoted by upper-case letters.}, $\bm{\Sigma} = \text{diag}\left(\sigma^2_1, \sigma^2_2,\ldots, \sigma^2_m\right)$ denotes the $m \times m$ diagonal matrix with residual variances, $\bm{0}_k$ denotes a $k$-dimensional vector with zeros, and $\bm{I}_k$ denotes the $k \times k$ identity matrix. Hence, conditional on the latent factors, the observations on the $m$ variables are assumed to be uncorrelated with each other. Marginally, however, the observations are usually not uncorrelated and they are distributed as
\begin{equation}
\bm{y}_t \sim \mathcal{N}_m\left(\bm{0}_m, \bm{\Omega}\right),
\end{equation}
where $\bm{\Omega} = \bm{\beta} \bm{\beta}^\top + \,\bm{\Sigma}$.

\begin{figure}[!tb]
 \begin{center}
	\includegraphics[width = .85\textwidth]{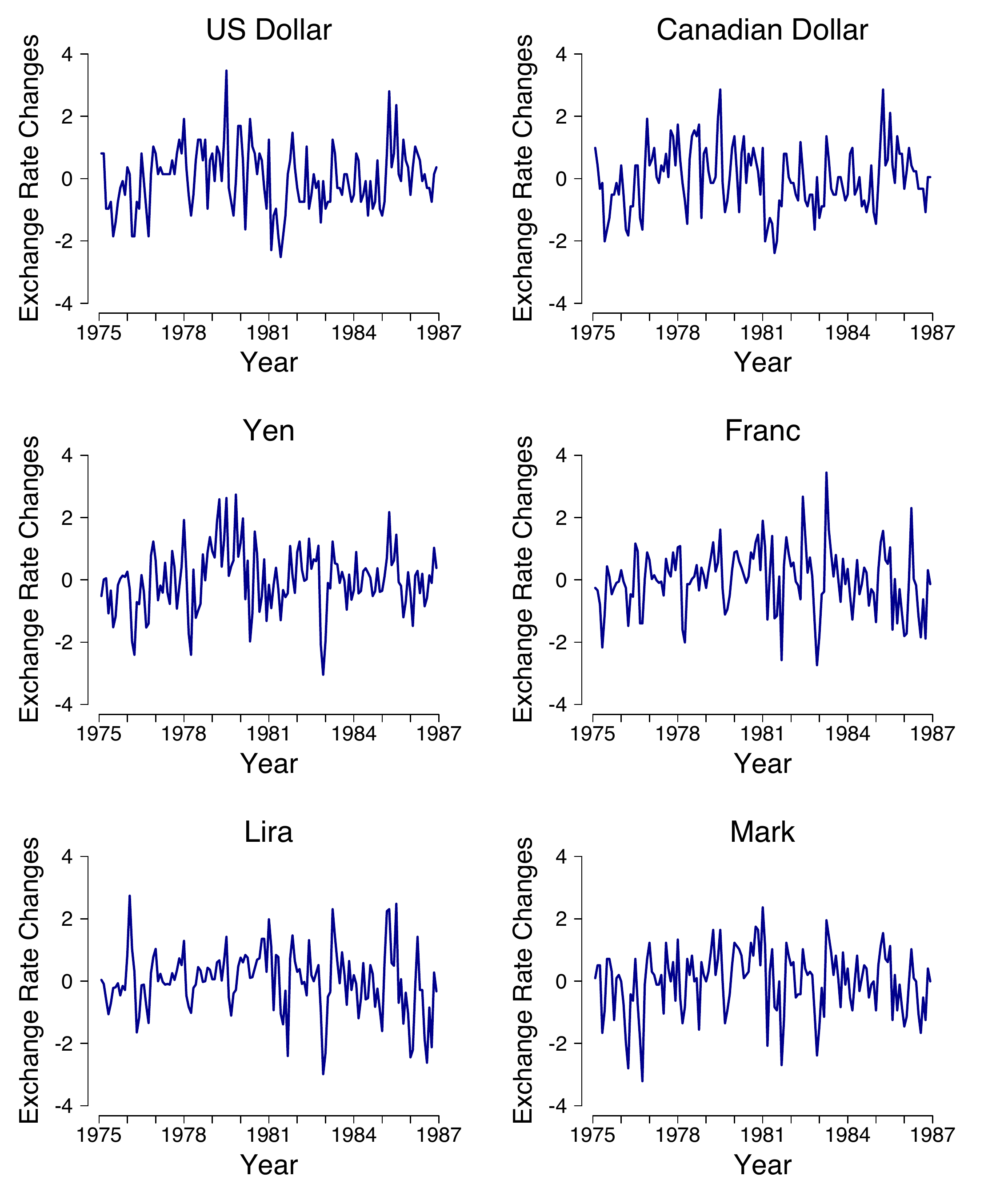}
	\caption{Changes in monthly international exchange rates for pounds sterling from January 1975 to December 1986 \citep[pp. 612 -- 615]{WestHarrison1997}. Currencies tracked are US Dollar (US), Canadian Dollar (CAN), Japanese Yen (JAP), French Franc (FRA), Italian Lira (ITA), and the (West) German Mark (GER). Each series has been standardized with respect to its sample mean and standard deviation. Figure reproduced from \citet{LopesWest2004}. Figure available at \url{https://tinyurl.com/ybtdddyv} under CC license \url{https://creativecommons.org/licenses/by/2.0/}.}
	\label{fig:ier}
	 \end{center}
\end{figure}
Here we reanalyze a data set that contains the changes in monthly international exchange rates for pounds sterling from January 1975 to December 1986 \citep[pp. 612-615]{WestHarrison1997}. Currencies tracked are US Dollar (US), Canadian Dollar (CAN), Japanese Yen (JAP), French Franc (FRA), Italian Lira (ITA), and the (West) German Mark (GER). Figure~\ref{fig:ier} displays the data.\footnote{Each series has been standardized with respect to its sample mean and standard deviation. These standardized data are included in the \pkg{bridgesampling} package.}  Using different computational methods, including bridge sampling, \citet{LopesWest2004} estimated the marginal likelihoods and posterior model probabilities for a factor model with one, two, and three factors. As before, this allows us to compare the results from the \pkg{bridgesampling} package to the results reported in the literature.
To identify the model, the factor loading matrix $\bm{\beta}$ is constrained to be lower-triangular \citep{LopesWest2004}. The diagonal elements of $\bm{\beta}$ are constrained to be positive by assigning them standard half-normal priors with lower truncation point zero: $\beta_{jj} \sim \mathcal{N}(0, 1)_{T(0,)}, \; j = 1,2,\ldots, k$, and the lower-diagonal elements are assigned standard normal priors. The residual variances are assigned inverse-gamma priors of the form $\sigma^2_i \sim \text{Inverse-Gamma}(\nu/2,\nu s^2/2), \; i = 1,2,\ldots, m$, where $\nu = 2.2$ and $\nu s^2 = 0.1$ \citep[for details, see][]{LopesWest2004}.

The first step in our reanalysis is to specify the \proglang{Stan} model as the character string \code{model\_code}.
We can then fit the three models corresponding to $k = 1$, $k = 2$, and $k = 3$ latent factors and estimate the log marginal likelihoods using \pkg{bridgesampling} as follows:\footnote{The complete code can be found in the supplemental material, on the Open Science Framework (\url{https://osf.io/3yc8q/}), and is also available at  \code{?ier}. Note that we specify initial values using a custom \code{init\_fun} function. This function may need to be changed for different applications. Furthermore, it is strongly advised to check that the chains have indeed converged since we sometimes encountered convergence issues with this model. Note that the results are dependent on the compiler and the optimization settings. Thus, even with identical seeds results can differ slightly from the ones reported here.}
\begin{Sinput}
R> library("rstan")
R> library("bridgesampling")
R> data("ier")
R> cores <- 4
R> options(mc.cores = cores)                    # for parallel MCMC chains
R> model <- stan_model(model_code = model_code) # compile model
R> set.seed(1)
R> stanfit <- bridge <- vector("list", 3)
R> for (k in 1:3) {
+   stanfit[[k]] <- sampling(model,
+                            data = list(Y = ier, T = nrow(ier),
+                                        m = ncol(ier), k = k),
+                            iter = 11000, warmup = 1000, chains = 4,
+                            init = init_fun(nchains = 4, k = k,
+                                            m = ncol(ier)),
+                            cores = cores, seed = 1)
+   bridge[[k]] <- bridge_sampler(stanfit[[k]], method = "warp3",
+                                 repetitions = 10, cores = cores)
+ }
\end{Sinput}
Note that in this example, we use the \code{"warp3"} method instead of the \code{"normal"} method. Furthermore, since the \code{error_measures} function cannot be used when the estimate has been obtained using \code{method = "warp3"} with \code{repetitions = 1}, we set \code{repetitions = 10} to obtain an empirical estimate of the estimation uncertainty (conditional on the posterior samples).
We also select parallel computation by setting \code{cores = 4}.
The \code{summary} method provides a convenient overview of the estimate and the estimation uncertainty.
For instance, for the 2-factor model, we obtain as output:
\begin{Sinput}
R> summary(bridge[[2]])
\end{Sinput}
\begin{Soutput}
Bridge sampling log marginal likelihood estimate 
(method = "warp3", repetitions = 10):

-903.4522

Error Measures:

Min: -903.4565
Max: -903.4481
Interquartile Range: 0.002682305

Note:
All error measures are based on 10 estimates.
\end{Soutput}

Table~\ref{tab:ier_logmls} displays for each of the three factor models (i.e., $k = 1$, $k = 2$, $k = 3$) the median log marginal likelihood (logml) across repetitions, the minimum/maximum log marginal likelihood across repetitions, and the log marginal likelihood value reported in \citet{LopesWest2004} based on bridge sampling. Note that the negative infinity reported by \citet{LopesWest2004} might be due to a numerical problem. For the 1-factor model and the 2-factor model, the log marginal likelihoods obtained via \pkg{bridgesampling} are very similar to the ones reported in \citet{LopesWest2004}. Furthermore, the narrow range of the estimates indicates that the estimation uncertainty is small (conditional on the posterior samples, as described above).

To examine the support for the three different models (i.e., different numbers of latent factors), we can use the \code{post_prob} function to compute posterior model probabilities. By default, the function assumes that all models are equally likely a priori; this can be adjusted using the \code{prior_prob} argument. Furthermore, the \code{model_names} argument can optionally be used to provide names for the models. Here we use the default of equal prior model probabilities and we obtain:
\begin{Sinput}
R> post_prob(bridge[[1]], bridge[[2]], bridge[[3]],
+            model_names = c("k = 1", "k = 2", "k = 3"))
\end{Sinput}
\begin{Soutput}
             k = 1     k = 2     k = 3
[1,]  6.278942e-49 0.8435919 0.1564081
[2,]  6.309963e-49 0.8491811 0.1508189
[3,]  6.373407e-49 0.8554668 0.1445332
[4,]  6.511718e-49 0.8739641 0.1260359
[5,]  6.582895e-49 0.8805172 0.1194828
[6,]  6.384273e-49 0.8596401 0.1403599
[7,]  6.469723e-49 0.8736989 0.1263011
[8,]  6.403270e-49 0.8616183 0.1383817
[9,]  6.426132e-49 0.8635907 0.1364093
[10,] 6.417346e-49 0.8592737 0.1407263
\end{Soutput}
Each row presents the posterior model probabilities based on one repetition of the bridge sampling procedure for all three models (i.e., each row sums to one). Hence, there are as many rows as \code{repetitions}.\footnote{Note that the output of the \code{post\_prob} function can be directly passed to the \code{boxplot} function which allows one to visualize the estimation uncertainty in the posterior model probabilities across repetitions.} The 2-factor model receives most support from the observed data. This is in line with \citet{LopesWest2004}, who also preferred the 2-factor model;\footnote{Note that \citet{LopesWest2004} report a posterior model probability of 1 for the 2-factor model. However, this estimate may be inflated by the infinite log marginal likelihood value for the 3-factor model.} based on the factor loadings, they proposed the presence of a North American factor and a European Union factor.
\begin{table}[t]
	\centering
		\normalsize
		\caption{Log marginal likelihood (logml) estimates for the $k = 1$, $k = 2$, and $k = 3$ factor model. The rightmost column displays the values based on bridge sampling reported in \citet{LopesWest2004}.}
	\begin{tabular}{crrrr}
		\toprule
		\textbf{Number of Factors} & \textbf{Median Logml} & \textbf{Min Logml} & \textbf{Max Logml} & \textbf{Lopes \& West} \\
        \hline
        $k = 1$ & -1014.271 & -1014.273 & -1014.269 & -1014.5 \\
        $k = 2$ & -903.452 & -903.457 & -903.448 & -903.7 \\
        $k = 3$ & -905.271 & -905.454 & -905.138 & $-\infty$ \\
		\bottomrule
        \label{tab:ier_logmls}
	\end{tabular}
\end{table}

\section{Discussion}

This paper introduced \pkg{bridgesampling}, an \proglang{R} package for computing marginal likelihoods, Bayes factors, posterior model probabilities, and normalizing constants in general.
We have demonstrated how researchers can use \pkg{bridgesampling} to conduct Bayesian model comparisons in a generic, user-friendly way: researchers need only provide posterior samples, a function that computes the log of the unnormalized posterior density, the data, and lower and upper bounds for the parameters.
Furthermore, we have described the \proglang{Stan} interface which makes it even easier to obtain the marginal likelihood: researchers need only provide a \code{stanfit} object and the \pkg{bridgesampling} package will automatically produce an estimate of the log marginal likelihood.\footnote{Similar to the \code{stanfit} method, the \code{bridge\_sampler} method for \pkg{nimble} only requires the fitted object (of class \code{MCMC\_refClass}) and extracts all necessary information for computing the marginal likelihood (including the function for computing the unnormalized log posterior density and the parameter bounds). However, at the time of writing we have not yet tested this method in the same intensity as the \code{stanfit} method. We will add a vignette describing the \pkg{nimble} interface in more detail when we have done so.}
In other words, the \pkg{bridgesampling} package makes it possible to obtain marginal likelihood estimates for any model that can be implemented in \proglang{Stan} (in a way that retains the constants).
By combining the \proglang{Stan} state-of-the-art No-U-Turn sampler with \pkg{bridgesampling}, researchers are provided with a general purpose, easy-to-use computational solution to the challenging task of comparing complex Bayesian models.

As practical advice, we recommend to keep the following four points in mind when using the \pkg{bridgesampling} package \citep[see also][]{GronauEtAlwarp3,GronauEtAlLBA}. First, one should always check the posterior samples carefully. A successful application of bridge sampling requires a sufficient number of representative samples from the posterior distribution. Thus, it is important to use efficient sampling algorithms and, in case of MCMC sampling, it is crucial that researchers confirm that the chains have converged to the joint posterior distribution.
In addition, researchers need to make sure that the model does not contain any discrete parameters since those are currently not supported. This may sound more restrictive than it is. In practice the solution is to marginalize out the discrete parameters, something that is often possible.
Note the similarity to \proglang{Stan} which also deals with discrete parameters by marginalizing them out \citep[section 15]{stanmanual}.
Furthermore, as demonstrated in the examples, for conducting model comparisons based on bridge sampling, the number of posterior samples often needs to be an order of magnitude larger than for estimation. This of course depends on a number of factors such as the complexity of the model of interest, the number of posterior samples that one usually uses for estimation, the posterior sampling algorithm used, and also the accuracy of the marginal likelihood estimate that one desires to achieve.

Second, one should always assess the uncertainty of the bridge sampling estimate.
In case the uncertainty is deemed too high, one can attempt to achieve a higher precision by increasing the number of posterior samples or, in case \code{method = "normal"}, by using the more sophisticated \code{method = "warp3"} instead (see the third point below). Users of the \pkg{bridgesampling} package have different options for assessing the estimation uncertainty. In our opinion, the ``gold standard'' may be to obtain an empirical uncertainty assessment by repeating the bridge sampling procedure multiple times, each time using a fresh set of posterior samples. This approach allows users to assess the uncertainty directly for the quantity of interest. For instance, if the focus is on computing a Bayes factor, users may repeat the following steps: (a) obtain posterior samples for both models, (b) use the \code{bridge\_sampler} function to estimate the log marginal likelihoods, (c) compute the Bayes factor using the \code{bf} function. The variability of these Bayes factor estimates across repetitions then provides an assessment of the uncertainty. For certain applications, this approach may be infeasible due to computational restrictions. If this is the case and \code{method = "normal"}, we recommend to use the approximate errors based on \citet{FruehwirthSchnatter2004} which are available through the \code{error\_measures} function. As mentioned before, we have found these approximate errors to work well for \code{method = "normal"}, but not for \code{method = "warp3"} which is the reason why they are not available for the latter method. Alternatively, one can also assess the estimation uncertainty by setting the \code{repetitions} argument to an integer larger than one. This provides an assessment of the estimation uncertainty due to variability in the samples from the proposal distribution, but it should be kept in mind that this does not take into account variability in the posterior samples.

Third, one should consider whether using the more time-consuming Warp-III method may be beneficial. The accuracy of the estimate is governed not only be the number of samples, but also by the overlap between the posterior and the proposal distribution \citep[e.g.,][]{MengWong1996,MengSchilling2002}. The \pkg{bridgesampling} package attempts to maximize this overlap by (a) focusing on one marginal likelihood at a time which allows one to use a convenient proposal distribution which closely resembles the posterior distribution, (b) using a proposal distribution which matches the mean vector and covariance matrix of the posterior samples (i.e., \code{method = "normal"}) or additionally also the skewness (i.e., \code{method = "warp3"}).
Consequently, as mentioned before, \code{method = "warp3"} will always be as precise or more precise than \code{method = "normal"}; however, it also takes about twice as long. We have found that in many applications, \code{method = "normal"} works well, however, in case the posterior is skewed (crucially, this refers to the joint posterior of the quantities that have been transformed to the real line), \code{method = "warp3"} may be the better choice. When in doubt, we believe that it may be beneficial to also explore the Warp-III results -- if this is computationally feasible -- to see how much (if any) improvement in precision is achieved by taking into account potential skewness.
It should be kept in mind that, in case the posterior distribution exhibits multiple modes, the overlap of the two distributions may still be subject to improvement -- even when using \code{method = "warp3"}. The development of efficient bridge sampling variants for these cases is subject to ongoing research \cite[e.g.,][]{WangMeng2016,FruehwirthSchnatter2004}.

Forth, users should carefully think about the choice of prior distribution.
Even though the \pkg{bridgesampling} package enables researchers to compute the marginal likelihoods in an almost black-box manner, this does not imply that the user can mindlessly exploit the package functionality to conduct Bayesian model comparisons. As is apparent from Equation~(\ref{eq:marginal_likelihood}), Bayesian model comparisons depend on the choice of the parameter prior distribution. Crucially, the prior distribution has a lasting influence on the results. Hence, meaningful Bayesian model comparisons require that researchers carefully consider their parameter prior distribution \citep[e.g.,][]{LeeVanpaemelinpress}, engage in sensitivity analyses, or use default prior choices that have certain desirable properties such as model selection and information consistency \citep[e.g.,][]{Jeffreys1961,BayarriEtAl2012,LyEtAl2016}.\footnote{Note that in the first example (i.e., the Bayesian $t$-test) we have used prior distributions which lead to these desirable properties. However, in the second and third example, we simply used the prior distributions that have been used in the literature so that we could compare our results to the reported results.}
 Thus, the \pkg{bridgesampling} package removes the computational hurdle of obtaining the marginal likelihood, thereby allowing researchers to spend more time and effort on the specification of meaningful prior distributions.

It should also be kept in mind that there may be cases in which the bridge sampling procedure may not be the ideal choice for conducting Bayesian model comparisons. For instance, when the models are nested it might be faster and easier to use the Savage-Dickey density ratio \citep{DickeyLientz1970,WagenmakersEtAl2010SDPsychologists}. Another example is when the comparison of interest concerns a very large model space, and a separate bridge sampling based computation of marginal likelihoods may take too much time. In this scenario, Reversible Jump MCMC \citep{Green1995} may be more appropriate. The downside of Reversible Jump MCMC is that it is usually problem-specific and cannot easily be applied in a generic fashion to different nested and non-nested model comparison problems \citep[but see][]{rjmcmc}. The goal with the \pkg{bridgesampling} package, however, was exactly that: to provide users with a generic way of computing marginal likelihoods which can in principle be applied to any Bayesian model comparison problem.

In the future, we hope that it may be possible to add \pkg{bridgesampling} support for a number of \proglang{R} packages, such as the \pkg{MCMCglmm} package \citep{Hadfield2010}, the \proglang{JAGS} interface of the \pkg{mgcv} package \citep{Wood2016}, the \pkg{glmmBUGS} package \citep{BrownZhou2018}, or the \pkg{blavaan}\footnote{Note that the \pkg{blavaan} package already provides approximate marginal likelihoods for the models that are obtained via a Laplace approximation.} package \citep{Merkle2018} so that users could conduct Bayesian model comparisons in a black box way similar to the \proglang{Stan} interface. For packages that use themselves \proglang{Stan} for fitting the models, adding \pkg{bridgesampling} support is relatively straightforward:  the only potential change that would have to be implemented is to make sure that the models are coded such that all constants are retained (as explained in section 4). Once this is achieved, computing the relevant quantities via \pkg{bridgesampling} works as described in the \proglang{Stan} examples. For packages that do not use \proglang{Stan} to fit the models, the main difficulty is specifying the unnormalized posterior density function and the parameter bounds in an automatized way. This is also the reason why there is currently no black box interface to \proglang{JAGS} since, to the best of our knowledge, specifying these quantities in an automatized way is not trivial. Nevertheless, if this hurdle could be overcome, adding \pkg{bridgesampling} support would be straightforward.

In sum, the \pkg{bridgesampling} package provides a generic, accurate, easy-to-use, automatic, and fast way of computing marginal likelihoods and conducting Bayesian model comparisons. With the computational challenge all but overcome, researchers can spend more time and effort on addressing the conceptual challenge that comes with Bayesian model comparisons: specifying prior distributions that are either robust or meaningful. 

\section{Acknowledgements}
This research was supported by a Netherlands Organisation for Scientific Research (NWO) grant to QFG (406.16.528), by an NWO Vici grant to EJW (016.Vici.170.083), and by the Berkeley Initiative for Transparency in the Social Sciences, a program of the Center for Effective Global Action (CEGA), with support from the Laura and John Arnold Foundation. 



\bibliography{references,referenties}

\end{document}